\documentclass[prb,twocolumn,showpacs]{revtex4}
\usepackage{graphicx}
\usepackage{amsfonts}
\usepackage{bbm}
\usepackage{color}

\newcommand{\ba}{\begin{eqnarray}}
\newcommand{\be}{\begin{equation}}
\newcommand{\ea}{\end{eqnarray}}
\newcommand{\ee}{\end{equation}}

\newcommand{\ignore}[1]{}

\begin{document}

\title{Memristive properties of single-molecule magnets}

\author{Carsten Timm}
\email{carsten.timm@tu-dresden.de}
\affiliation{Institute for Theoretical Physics, Technische Universit\"at
Dresden, 01062 Dresden, Germany}

\author{Massimiliano Di Ventra}
\email{diventra@physics.ucsd.edu}
\affiliation{Department of Physics, University of California, San Diego, La
Jolla, California 92093, USA}

\date{July 23, 2012}

\begin{abstract}
Single-molecule magnets weakly coupled to two ferromagnetic leads act as
memory devices in electronic circuits---their response depends on history, not
just on the instantaneous applied voltage. We show that magnetic anisotropy
introduces a wide separation of time scales between fast and slow relaxation
processes in the system, which leads to a pronounced memory dependence in a wide
intermediate time regime. We study the response to a harmonically varying bias
voltage from slow to rapid driving within a master-equation approach.
The system is not purely memristive but shows a partially capacitive response
on short time scales. In the intermediate time regime, the molecular spin can be
used as the state variable in a two-terminal molecular memory device.
\end{abstract}


\maketitle

\section{Introduction}

Electronic transport through magnetic molecules has been extensively studied
both experimentally\cite{PPG02,LSB02,HGF06,JGB06,JRB07,GTT08,
IRH08,SKC09,ZBO10,RVE11,KIL11,MKR11,MGD12} and
theoretically.\cite{ElT05,RWS06,LeM06,TiE06,ElT06,
LeL07,MiB07a,GoL07,CUB07,JIM08,RLW09,LZS09,MWB09,ElT10,SoC10,JCG10,Yu10,CRA11,
MWB11,LEK12,BAL12} The magnetic moment of these molecules can be realized by
one or more transition-metal ions with organic ligands, by organic radicals, or
by endohedral fullerenes. Transition-metal ions in a molecular ligand field
typically show sizable magnetic anisotropy. In the case of a pronounced easy
axis, one then speaks of \emph{single-molecule magnets} (SMMs). For
memory devices, easy-axis anisotropy is desirable since it introduces an energy
barrier for spin reversal and thereby stabilizes the spin in the up or down
orientation.\cite{TiE06}

Driving the system with an external electric field allows the
control and manipulation of the molecular spin state, and consequently
writing and reading of information using SMMs.\cite{TiE06}
Clearly, the response of the system to the applied electric field
is not purely instantaneous---the system has memory.
This can be compared to the response of SMMs to a magnetic
field. At low temperatures, they show pronounced hysteresis in the
magnetization (of bulk crystals of noninteracting SMMs) vs the
magnetic field.\cite{SGC93,FrS10}

Similarly, if a harmonically varying bias voltage is applied, we expect
hysteresis loops in the spin, charge, and current dynamics vs the
applied voltage to develop, whose amplitude depends on the voltage
amplitude and frequency---as in other systems whose spin
polarization can be controlled by a voltage or current; see, e.g.,
Ref.~\onlinecite{pershin08a}.
At very low frequencies, the spin dynamics can easily follow the external field
and little or no hysteresis is expected. At very high frequencies, the spin
dynamics are ``frozen.''
There is, however, an intermediate frequency range---comparable to
the inverse spin-relaxation time(s)---where the hysteresis is most pronounced.
In this range, the device state, at any given time, is strongly dependent on the
history of states through which the system has evolved. In that
case, we expect the resistance of the device to be a function of the state
variable $x$ that describes its memory (the spin
polarization) and possibly of the protocol with which the voltage $V(t)$
has been applied, i.e., the entire wave form of $V(t)$. In other
words, the resistive response can be characterized by a function of the type
$R(x,V,t)$. Such a device goes under the name of memristive (for ``memory
resistive'') system.\cite{chua76a,diventra09a,PeV11} Resistors with memory are
experiencing a surge of {research activity, in part due to promising
applications} in memory storage, but also because of
their ubiquity in diverse areas ranging from nontraditional computing to
biophysics.\cite{pershin09b,pershin10c,pershin12a} It is natural to think
that SMMs form another example of
memristive systems, with the molecular spin playing the role of internal state
variable. This would have the added advantage of combining memristive and
spintronics functionality\cite{pershin08a,wang2009} in molecular junctions.
Indeed, Miyamachi \textit{et al.}\cite{MGD12} have recently
demonstrated memristive behavior of single
$\mathrm{Fe}(1,10-\mathrm{phenanthroline})_2(\mathrm{NCS})_2$ molecules
on CuN under a scanning tunneling microscope. In their case, the molecule is
switched between a high-spin ($S=2$) and a low-spin ($S=0$) state of the
$\mathrm{Fe}^{2+}$ ion, which is connected with a change in conformation and
conductance. Our case is quite different: We consider the switching of a
molecular local spin of fixed length $S$ over an anisotropy barrier.

In this paper, we show that the response of this system is only partially
memristive. In addition, capacitive components emerge
on short time scales. The capacitive components are related to the
charging energy of the molecule and thus to the Coulomb repulsion of
electrons.

Transport through magnetic molecules is typically dominated by a strong exchange
interaction between the spin of mobile electrons and the local molecular spin,
in addition to the large Coulomb repulsion.
We are thus faced with a strongly interacting
nonequilibrium system, which makes a quantitative description difficult in
general. However, if the coupling between the molecule and the metallic leads
is weak, as is often the case in break junctions, this coupling can be used as
the small pa\-ra\-me\-ter in a perturbative approach. This can be done in the
framework of the master equation, which has the advantage that the strong
interactions within the molecule can be treated exactly. The master equation has
been applied to transport through magnetic molecules by various
groups.\cite{ElT05,RWS06,TiE06,ElT06,LeL07,MiB07a,GoL07,JIM08,RLW09,LZS09,
MWB09,ElT10,SoC10} It provides an ensemble description, not a description of
individual time series of single-molecule devices. Statements we make about the
memristive properties are thus to be understood in an ensemble sense. On the
other hand, devices consisting of a {monolayer of weakly interacting molecules}
between metallic electrodes, in which many molecules conduct in parallel, are
self-averaging. In this case, we predict the time-dependent observables of a
single device.

The analysis of SMMs as memory devices requires us to study
their dynamics under a time-dependent bias. In Refs.\ \onlinecite{TiE06} and
\onlinecite{ElT06}, their relaxation for constant or
suddenly switched bias has been considered. Here, we consider the
current, charge, and spin response to a harmonically varying bias
$V(t) = V_0\, \sin \omega t$, which is easily realizable experimentally. For a
non-magnetic molecule involving a
vibrational mode, the response to a harmonic voltage has been recently
studied by Donarini \textit{et al.}\cite{DYG12} In this case, Franck-Condon
blockade leads to interesting dynamics.\cite{Koch,DYG12}

The remainder of this paper is organized as follows: In Sec.\ \ref{sec.model}
we present our model, followed by a discussion of the master-equation approach
in Sec.\ \ref{sec.master}. The results are presented and discussed in Sec.\
\ref{sec.results}. Finally, in Sec.\ \ref{sec.summary} we summarize the main
points, address possible limiting effects, and draw some conclusions.

\section{Model}
\label{sec.model}

Our device consists of a magnetic molecule coupled to two ferromagnetic leads.
The full system is described by the Hamiltonian $H=H_\mathrm{mol} +
H_\mathrm{leads} + H_\mathrm{hyb}$, where the molecular Hamiltonian
reads\cite{TiE06,ElT06,MiB07a}
\be
H_\mathrm{mol} = \epsilon_d \sum_\sigma d^\dagger_\sigma
  d_\sigma + U d^\dagger_\uparrow d_\uparrow d^\dagger_\downarrow
  d_\downarrow - J \mathbf{s}\cdot\mathbf{S} - K_2\, (S^z)^2 ,
\ee
where $d^\dagger_\sigma$ ($d_\sigma$) creates (annihilates) an electron of
spin $\sigma$ in the molecular orbital with energy $\epsilon_d$, $\mathbf{s}
\equiv \sum_{\sigma\sigma'}
d^\dagger_\sigma (\mbox{\boldmath$\sigma$}_{\sigma\sigma'}/2) d_\sigma$ is the
corresponding spin operator, and $\mathbf{S}$ is the spin operator of a local
spin of length $S$. The two spins are coupled by the exchange interaction $J$
and the local spin is subject to an easy-axis anisotropy of strength $K_2>0$.
The extension of this model by including more than one electronic orbital, or
more than
one local spin, does not pose any conceptual difficulties. In
a break-junction setup, the on-site energy $\epsilon_d$ could be tuned by a gate
voltage. However, this tunability is not necessary for our conclusions.

\begin{figure}[tbh]
\includegraphics[width=3.00in,clip]{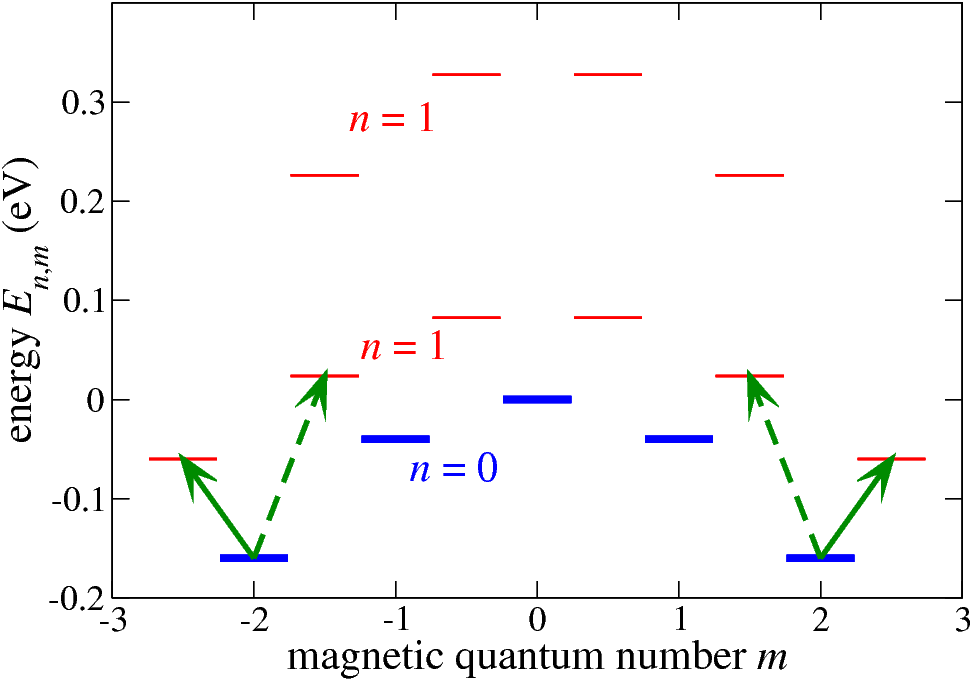}
\caption{\label{fig.Evsm}(Color online) Molecular energy levels $E_{n,m}$ vs
magnetic
quantum number $m$ for $\epsilon_d = 0.2\,\mathrm{eV}$, $U=10\,\mathrm{eV}$,
$J=0.1\,\mathrm{eV}$, $K_2=40\,\mathrm{meV}$, and $S=2$. Levels for electron
occupation number $n=0$ ($n=1$) are shown as heavy blue (medium red) bars.
Levels for electron number $n=2$ are outside the range of the plot. The arrows
indicate the lowest-energy sequential-tunneling transitions out of the
ground states.}
\end{figure}

The molecular Hamiltonian $H_\mathrm{mol}$ commutes with the \textit{z}
component of the total spin
$\mathbf{S}_\mathrm{tot}\equiv\mathbf{s}+\mathbf{S}$ so that the eigenvalue $m$
of $S_\mathrm{tot}^z$ is a good quantum number. We show in
Fig.\ \ref{fig.Evsm} the energy levels of $H_\mathrm{mol}$ vs $m$ for the
parameter values $\epsilon_d = 0.2\,\mathrm{eV}$, $U=10\,\mathrm{eV}$,
$J=0.1\,\mathrm{eV}$, $K_2=40\,\mathrm{meV}$, and $S=2$, which we will also use
below to illustrate our results. Note that we are using a large
value of the anisotropy energy in order
to most clearly show the generic behavior. For the smaller anisotropies of
SMMs,\cite{FrS10} the
interesting physics would occur in a narrower voltage range and at lower
temperatures. For $n=0$ and for $n=2$ electrons on the molecule, the
total spin is just given by the local spin $\mathbf{S}$ so that there are $2S+1$
levels in these charge sectors. For one electron, $n=1$, its spin $1/2$ combines
with the local spin $S$ to form two multiplets of $2(S-1/2)+1=2S$ and
$2(S+1/2)+1=2S+2$ states. The splitting between the two multiplets is on the
order of $JS$. The easy-axis anisotropy leads to the parabolic dispersion seen
for all multiplets in Fig.\ \ref{fig.Evsm}. The total dimension of the molecular
Fock space is $N_F=4(2S+1)$.

The leads are described by the Hamiltonian
\be
H_\mathrm{leads} = \sum_{\nu\mathbf{k}\sigma} \epsilon_{\nu\mathbf{k}\sigma}\,
  c^\dagger_{\nu\mathbf{k}\sigma} c_{\nu\mathbf{k}\sigma} ,
\ee
where $c^\dagger_{\nu\mathbf{k}\sigma}$ ($c_{\nu\mathbf{k}\sigma}$) creates
(annihilates) an electron with wave vector $\mathbf{k}$ and spin $\sigma$ in
lead $\nu=L,R$. Below, we will assume that the leads are ferromagnetic with
opposite magnetizations parallel to the \textit{z} direction. The molecule and
the leads are coupled by the hybridization term
\be
H_\mathrm{hyb} = - \frac{t_\mathrm{hyb}}{\sqrt{N}} \sum_{\nu\mathbf{k}\sigma}
(c^\dagger_{\nu\mathbf{k}\sigma}
  d_\sigma + d^\dagger_\sigma c_{\nu\mathbf{k}\sigma}) ,
\ee
where, for simplicity, we have assumed the tunneling matrix
element $t_\mathrm{hyb}$ to be real
and independent of the lead index, wave vector, and spin. The
number of sites, $N$, in each lead is introduced by the Fourier
transformation into momentum space and drops out of the physical results.

\section{Master equation}
\label{sec.master}

The master-equation approach starts from the exact von Neumann equation (we set
$\hbar=1$),
\be
\frac{d\rho}{dt} = -i\, [H,\rho]
\ee
for the density (statistical) operator $\rho$ for the complete system. The
reduced density operator of the molecule is obtained by tracing
out the degrees of freedom of the leads, $\rho_\mathrm{mol} =
\mathrm{Tr}_\mathrm{leads}\,\rho$. The resulting equation of motion is
the master equation
for $\rho_\mathrm{mol}$.\cite{Nak58,Zwa60,ToM76,STH77,Ahn94,BrP02,Tim11} In
principle, an exact master equation that is local in time,
\be
\frac{d}{dt}\,\rho_\mathrm{mol} = -i\,[H_\mathrm{mol}(t),\rho_\mathrm{mol}(t)]
  - \mathcal{R}(t,t_0)\,\rho_\mathrm{mol}(t) ,
\label{3.TCLME.1}
\ee
can be derived even for time-dependent Hamiltonians if the molecule and the
leads were in a product state at some initial time $t_0$.\cite{Ahn94} The
first term on the right-hand side of Eq.\
(\ref{3.TCLME.1}) describes the time evolution of the decoupled molecule,
whereas the second term involving a linear superoperator $\mathcal{R}$ describes
relaxation.

In practice, approximations are needed to obtain superoperators $\mathcal{R}$
that preserve the positivity of the density operator. Here we
make three simplifications: (i) We employ the sequential-tunneling
approximation, which consists of keeping only terms up to second order in the
tunneling matrix element $t_\mathrm{hyb}$. This
approximation is reasonable if $\Gamma\ll k_B\Theta$, where $\Theta$ is
the temperature. (ii) We only consider reduced density operators
$\rho_\mathrm{mol}$ that are diagonal in the eigenbasis of $H_\mathrm{mol}$.
Actually, this is not an approximation: Below, we will always assume that the
time evolution starts from equilibrium or from a pure diagonal state. If we
start from such a diagonal $\rho_\mathrm{mol}$, the
exact time evolution does not generate nonzero off-diagonal components. This is
because off-diagonal components would correspond to superpositions (coherences)
of states with different charge or different spin component $S_\mathrm{tot}^z$.
The former would break $\mathrm{U}(1)$ charge symmetry, and could only be
expected in superconducting systems. The latter would lead to nonzero averages
of $S_\mathrm{tot}^x$ or $S_\mathrm{tot}^y$, which would require spontaneous
breaking of the spin-rotation symmetry around the \textit{z} axis.
(iii) We employ the Markov approximation, which posits that correlation
functions describing the memory of the leads decay on a time scale
$\tau_\mathrm{leads}$ that is much shorter than all experimentally relevant
time scales.
This requires the oscillation period $T$ of the bias voltage to satisfy
$T\gg \tau_\mathrm{leads}$. Since the leads are metals with typical relaxation
times in the femtosecond range, this is easily fulfilled.
The Markov approximation is necessary here since it allows us to use the
\emph{instantaneous} value of the bias voltage in the master equation.

The derivation of the resulting master equation is
standard\cite{Koch,MAM04,ElT05,TiE06,Tim08} and we only present the
results. We assume the bias voltage $V$ to be split
evenly between the two molecule-lead contacts. The two leads are assumed
to be identical ferromagnetic metals with their magnetizations along the
\textit{z} direction but of opposite sign. The relevant parameter is the ratio
$p = N_\mathrm{min}/N_\mathrm{maj}$ between the densities of states (assumed
to be energy independent) for minority-spin and majority-spin electrons.
We write the diagonal reduced density operator as
$\rho_\mathrm{mol}=\mathrm{diag}(P_1,P_2,\ldots)$, where the $P_i$ are the
probabilities of many-particle eigenstates $|i\rangle$ of $H_\mathrm{mol}$. The
master equation then takes the form
\be
\frac{dP_i}{dt} = \sum_j ( R_{j\to i}\, P_j - R_{i\to j}\, P_i ) ,
\label{3.Pauli.1}
\ee
with $R_{j\to i}$ the transition rate from state $|j\rangle$ to state
$|i\rangle$ due to sequential tunneling. The rate can be written as a sum over
contributions from spin up and spin down and from the left and right leads,
\be
R_{j\to i} = \sum_{\sigma=\uparrow,\downarrow} \, \sum_{\nu=L,R}
  R^{\sigma\nu}_{j\to i} ,
\ee
with\cite{TiE06}
\ba
R^{\uparrow L}_{j\to i} & = & \Gamma\, \bigg[ f\left(E_i - E_j +
  \frac{eV}{2}\right)\, |D^\uparrow_{ji}|^2 \nonumber \\
&& {}+ f\left(E_i - E_j - \frac{eV}{2}\right)\, |D^\uparrow_{ij}|^2 \bigg] ,
\label{3.rate.1} \\
R^{\downarrow L}_{j\to i} & = & p\,\Gamma\, \bigg[ f\left(E_i - E_j +
  \frac{eV}{2}\right)\, |D^\downarrow_{ji}|^2 \nonumber \\
&& {}+ f\left(E_i - E_j - \frac{eV}{2}\right)\, |D^\downarrow_{ij}|^2 \bigg] ,
\label{3.rate.2}  \\
R^{\uparrow R}_{j\to i} & = & p\,\Gamma\, \bigg[ f\left(E_i - E_j -
  \frac{eV}{2}\right)\, |D^\uparrow_{ji}|^2 \nonumber \\
&& {}+ f\left(E_i - E_j + \frac{eV}{2}\right)\, |D^\uparrow_{ij}|^2 \bigg] ,
\label{3.rate.3} \\
R^{\downarrow R}_{j\to i} & = & \Gamma\, \bigg[ f\left(E_i - E_j -
  \frac{eV}{2}\right)\, |D^\downarrow_{ji}|^2 \nonumber \\
&& {}+ f\left(E_i - E_j + \frac{eV}{2}\right)\, |D^\downarrow_{ij}|^2 \bigg] ,
\label{3.rate.4}
\ea
where $f(E)$ is the Fermi-Dirac distribution function, $E_i$ is the eigenenergy
of molecular state $|i\rangle$, $D^\sigma_{ij} \equiv \langle
i|d^\sigma|j\rangle$ are matrix elements of the electron annihilation operator
between molecular eigenstates, and
$\Gamma \equiv 2\pi|t_\mathrm{hyb}|^2 N_\mathrm{maj}$ quantifies the
coupling of majority electrons to the leads.

The average occupation number, the \textit{z} component of the electron spin,
and the \textit{z} component of the local spin are given by
\ba
\langle n\rangle & = &
  \sum_i P_i \,\langle i| \sum_\sigma d^\dagger_\sigma d_\sigma |i\rangle,
  \quad \\
\langle s^z \rangle & = &
  \sum_i P_i \,\langle i|\, \frac{d^\dagger_\uparrow d_\uparrow
  - d^\dagger_\downarrow d_\downarrow}{2}\, |i\rangle , \\
\langle S^z \rangle & = &
  \sum_i P_i \,\langle i| S^z |i\rangle ,
\ea
respectively. The average charge current between lead $\nu$ and the molecule
is\cite{TiE06}
\be
\langle I_\nu \rangle = -e\, \nu \sum_{ij} P_j\,
  (n_i - n_j) \, R^{\sigma\nu}_{j\to i} ,
\label{3.current.3}
\ee
where the numerical value of $\nu$ is $+1$ ($-1$) for the left (right) lead
and $n_i \equiv \langle i|\sum_\sigma d^\dagger_\sigma d_\sigma |i\rangle$.
The current is counted as positive if it is flowing from left to right.

While for the stationary state the left and right currents are equal, this is
not generally the case for time-dependent $\rho_\mathrm{mol}$.\cite{Mbook2008}
It is
then crucial to realize that an ammeter in, say, the left lead nevertheless
measures the \emph{symmetrized} current,
\be
I \equiv \frac{\langle I_L \rangle + \langle I_R \rangle}{2} .
\label{3.current.4}
\ee
The reason is the following: $\langle I_\nu \rangle$ only contains
the tunneling (particle) current through the contact between the molecule and
lead $\nu$. In addition, there are displacement currents across the contacts
resulting from charging of the molecule-lead capacitors.\cite{AnD95,WWG99} The
displacement currents are equal to the charging currents, as seen from the
simple example of a pure capacitor. An ammeter placed in the left lead picks up
the sum of the tunneling and the charging currents. By recalling that for a
symmetric
device the displacement currents in both barriers are equal in magnitude but
opposite in sign,\cite{XWW10} and that the sum of particle and displacement
current is divergence free, one can show that the sum of the particle and
displacement currents in the left contact, and thus the current in the ammeter,
equals the symmetrized particle current.\cite{XWW10}

The master equation (\ref{3.Pauli.1}) is solved numerically by discretizing time
and propagating the probabilities $P_i(t)$ forward step by step.
We will compare our results to the stationary solution at
constant bias voltage, which constitutes the limit of infinitely
slow driving. The stationary solution $P_i^\infty$ is found by setting the
left-hand side of Eq.\ (\ref{3.Pauli.1}) to zero, resulting in
\be
0 = \sum_j A_{ij}\, P^\infty_j
\ee
with
\be
A_{ij} = \left\{\begin{array}{ll}
  R_{j\to i} & \mbox{for $i\neq j$,} \\[1ex]
  - \sum_{k\neq i} R_{i\to k} & \mbox{for $i=j$.}
  \end{array}\right.
\ee
The stationary probability vector $\mathbf{P}^\infty =
(P_1^\infty,P_2^\infty,\ldots)$ is thus the right eigenvector of the
transition-rate matrix $A$ with zero eigenvalue. $A$ has at least one
vanishing eigenvalue, i.e., one stationary state, since $(1,1,\ldots,1)$ is
clearly a left eigenvector with zero eigenvalue.

We can also conclude that since our model is
ergodic---any state $|i\rangle$ can be reached from any other
state by a finite number of sequential-tunneling transitions with non-vanishing
rates---the stationary state is unique. However, the matrix $A$ is often
\emph{ill conditioned} since the rates $R_{j\to i}$ are spread over many orders
of magnitude due to the Fermi functions. This leads to the problem that
diagonalization routines working with machine precision obtain more than one
eigenvalue that is numerically zero, and thereby fail to find the unique
stationary state. We overcome this difficulty by
diagonalizing $A$ with high precision in {\sc mathematica}.\cite{Math} We use
$LU$ decomposition to estimate the $L^\infty$ condition number and adapt the
number of digits kept in the diagonalization so that the resulting
$\mathbf{P}^\infty$ contains at least $12$ significant digits.

For periodic bias voltage $V(t)$, the system will not relax toward a stationary
state but will approach a periodic cycle. We define the time-evolution
matrix $\Pi$ for one full period by
$\mathbf{P}(t+T) = \Pi\,\mathbf{P}(t)$.
To make $\Pi$ unique, we choose the time $t$ in such a way that
$V(t)=0$ and $V'(t)>0$, i.e., at vanishing bias voltage on the upsweep.
$\Pi$ is the stochastic matrix of a discrete-time Markov process. The periodic
state is characterized by the probability vector $\mathbf{P}^\mathrm{per} =
(P_1^\mathrm{per},P_2^\mathrm{per},\ldots)$ which is a right eigenvector of
the stochastic matrix with eigenvalue one,
$\mathbf{P}^\mathrm{per} = \Pi\, \mathbf{P}^\mathrm{per}$.
The stochastic matrix $\Pi$ has at least one unity eigenvalue, and this
eigenvalue is nondegenerate, since the discrete-time process is ergodic, in
analogy to the discussion for $A$ above.
Starting from $\mathbf{P}^\mathrm{per}$, the periodic time-dependence can
be obtained by integrating the master equation (\ref{3.Pauli.1}) over one
period. The stochastic matrix $\Pi$ is evaluated numerically by discretizing
time as
$\Pi = \prod_{0\le t< T} [ \mathbbm{1} + \Delta t\, A(t)]$,
where the transition-rate matrix $A$ now depends on time through the bias
voltage. We normalize $\Pi$ after each matrix multiplication such that $\sum_m
\Pi_{mn}=1$ for all $n$, which is required to conserve probability. $\Pi$
can also be ill conditioned and we apply the method sketched above to find the
unique periodic state.

\section{Results}
\label{sec.results}

\begin{figure}[tbh]
\includegraphics[width=3.10in,clip]{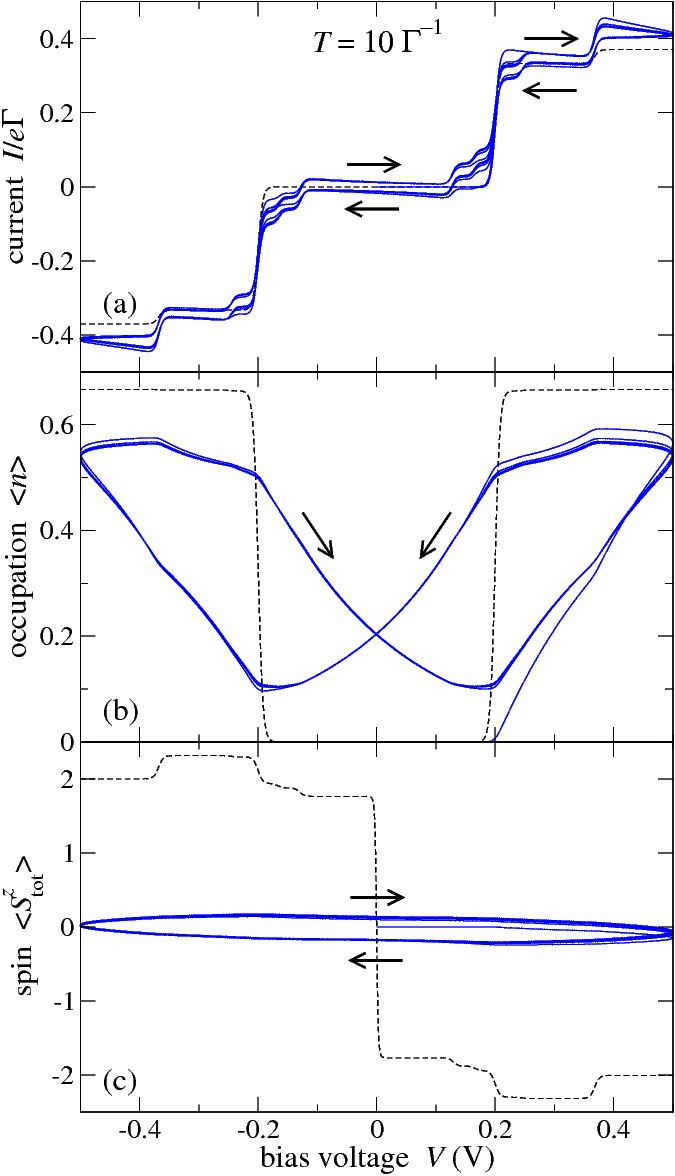}
\caption{\label{fig.onecase}(Color online) (a) Current, (b) occupation number,
and (c) \textit{z} component of the total spin vs bias voltage for the
SMM driven by a harmonic bias voltage of amplitude
$V_0=0.5\,\mathrm{V}$ and period $T=10\,\Gamma^{-1}$. Ten full periods are shown
as solid blue (gray) curves. The arrows indicate
the directions in which the loops are traversed. The black dashed curves refer
to the stationary state, corresponding to $T\to\infty$.
The spin polarization in the leads is
$p = N_\mathrm{min}/N_\mathrm{maj} = 0.5$. The other model parameters are
$\epsilon_d = 0.2\,\mathrm{eV}$, $U=10\,\mathrm{eV}$, $J=0.1\,\mathrm{eV}$,
$K_2=40\,\mathrm{meV}$, and $S=2$, as in Fig.\ \ref{fig.Evsm}.}
\end{figure}

Before analyzing the dependence on the period and the amplitude of the bias
voltage in detail, let us show the typical behavior of our system for one
parameter set. Figure \ref{fig.onecase} shows the approach to the periodic
regime when the system is initialized in its equilibrium state at time $t=0$ and
then driven by a bias $V(t) = V_0\, \sin \omega t$ with period
$T=2\pi/\omega$. The current, occupation number, and the \textit{z} component
of the total spin approach periodic behavior within a few periods. As
expected, all three
observables show hysteresis. Since the state of the system evidently depends on
its history, we immediately see that our system is indeed a memory
device.\cite{PeV11} Its internal state
is described by the probability vector $\mathbf{P}(t)$, which contains
$N_F-1$ independent real variables, where $N_F$ is the dimension of the
molecular Fock space. However, it is not a purely \emph{memristive} system: A
memristive system would satisfy equations of the form\cite{PeV11}
\ba
I(t) & = & G\big(V(t),\mathbf{P}(t)\big)\, V(t) ,
\label{4.memrist.2} \\
\frac{d\mathbf{P}}{dt} & = & \mathbf{F}\big(V(t),\mathbf{P}(t)\big) .
\ea
In our case, the second equation is the master equation
(\ref{3.Pauli.1}). On the other hand, our Eqs.\ (\ref{3.current.3}) and
(\ref{3.current.4}) for
the current cannot be written in the form of Eq.\
(\ref{4.memrist.2}) for all $V$. Equation (\ref{4.memrist.2}) implies
that the current vanishes for zero bias for a memristive system, whereas Fig.\
\ref{fig.onecase} shows that it does not vanish for our device---the hysteresis
loop is not pinched at $V=0$. Physically, this is because we have additional
capacitive or inductive effects. The
\textit{I}-\textit{V} characteristics of a harmonically driven pure capacitor
(inductor) show an ellipse that is traversed in the clockwise (counterclockwise)
direction. Since the loop in Fig.\ \ref{fig.onecase}(a) is clockwise, the
behavior is partially \emph{capacitive}. This is reasonable since the
molecule is transiently charged, as shown in Fig.\ \ref{fig.onecase}(b).

For comparison, Fig.\ \ref{fig.onecase} also shows the voltage dependences of
all observables in the stationary state (dashed curves). Note that the average
spin in the stationary state is nonzero and
depends on the bias voltage solely because the leads are magnetically polarized.
If electrons are predominantly moving from left to right ($I<0$),
the left lead injects predominantly spin-up electrons, while the right lead
absorbs predominantly spin-down electrons. The result is a positive spin
polarization on the molecule, as seen in Fig.\ \ref{fig.onecase}(c).
The stationary curves all show plateaus separated by thermally
broadened steps. The dynamical current and charge mainly relax toward the
stationary values at the plateaus but cannot follow the rapid changes at the
steps. The visibility of the relaxation suggests that the driving period is
comparable to the relevant relaxation times. On the other hand, the spin
hysteresis loop bears no resemblance to the stationary curve, showing that the
spin cannot relax rapidly enough to approach its stationary value. We will
discuss these points in what follows.

\subsection{Dependence on the voltage period}

\begin{figure}[tbh]
\includegraphics[width=3.10in,clip]{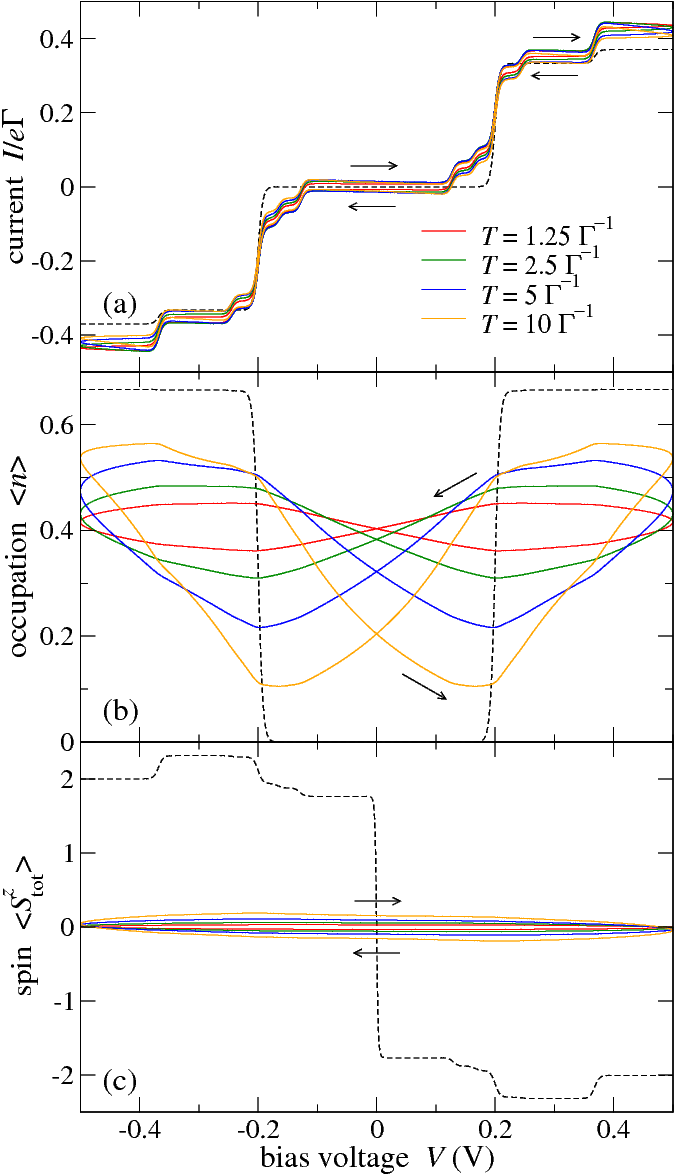}
\caption{\label{fig.per_fast}(Color online) (a) Current, (b) occupation number,
and (c) \textit{z} component of the total spin vs voltage for
voltages of amplitude $V_0=0.5\,\mathrm{V}$ and relatively short periods
$T=1.25$, $2.5$, $5$, and $10\,\Gamma^{-1}$. The other parameters are as in
Fig.\ \ref{fig.onecase}. A single hysteresis loop in the periodic regime is
shown. The arrows indicate the directions in which the loops are traversed.
The black dashed curves show the voltage dependence in the
stationary state.}
\end{figure}

\begin{figure}[tbh]
\includegraphics[width=3.10in,clip]{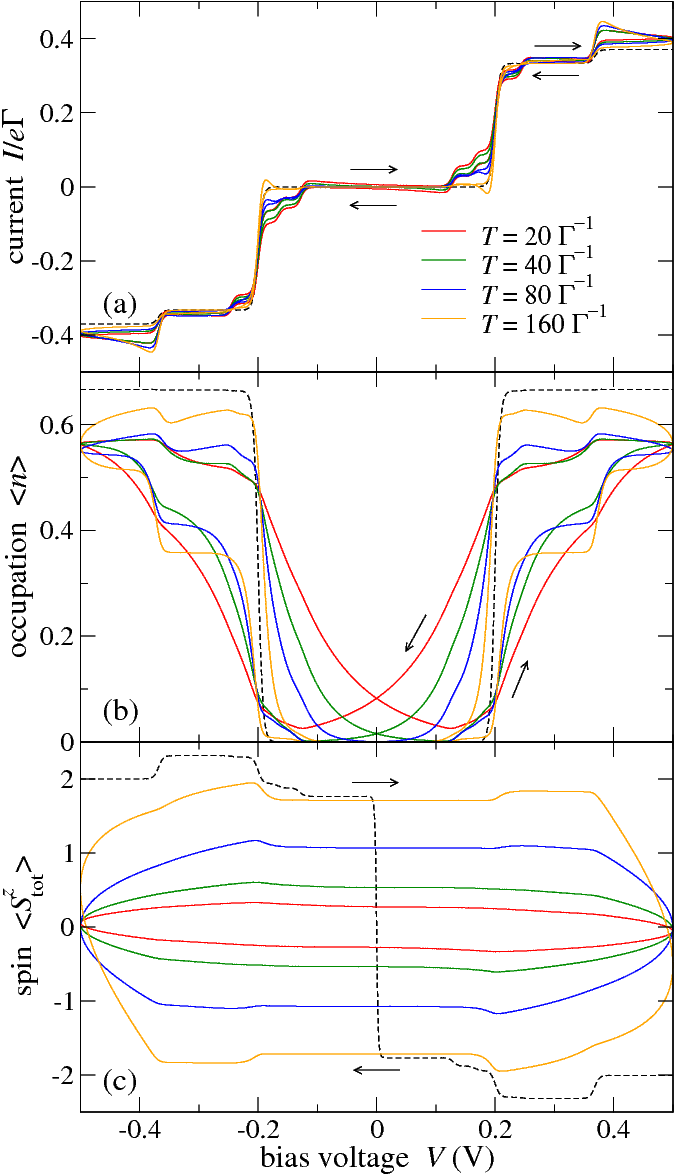}
\caption{\label{fig.per_medium}(Color online) (a) Current, (b) occupation
number, and (c) \textit{z} component of the total spin vs voltage for
voltages of amplitude $V_0=0.5\,\mathrm{V}$ and intermediate periods
$T=20$, $40$, $80$, and $160\,\Gamma^{-1}$. The other parameters are as in Fig.\
\ref{fig.onecase}. A single hysteresis loop in the periodic regime is shown.
The arrows indicate the directions in which the loops are traversed.
Black dashed curves: Stationary state.}
\end{figure}

\begin{figure}[tbh]
\includegraphics[width=3.10in,clip]{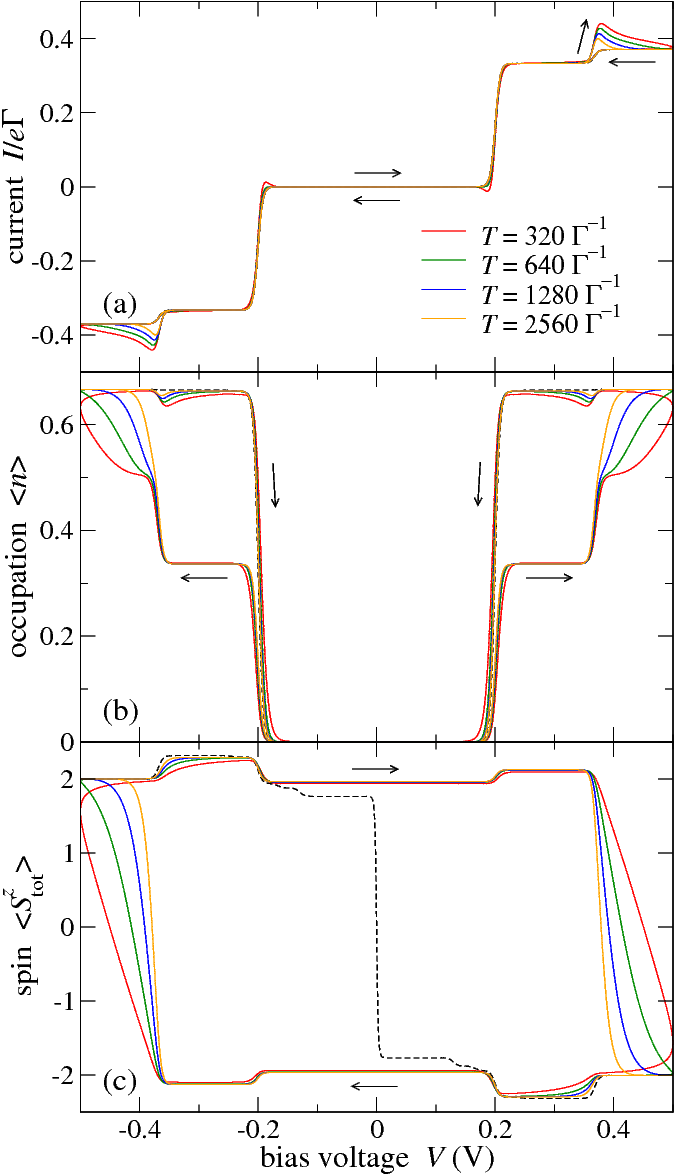}
\caption{\label{fig.per_slow}(Color online) (a) Current, (b) occupation number,
and (c) \textit{z} component of the total spin vs voltage for
voltages of amplitude $V_0=0.5\,\mathrm{V}$ and relatively long periods
$T=320$, $640$, $1280$, and $2560\,\Gamma^{-1}$. The other parameters are as in
Fig.\ \ref{fig.onecase}. A single hysteresis loop in the periodic regime is
shown. The arrows indicate the directions in which the loops are traversed.
Black dashed curves: Stationary state.}
\end{figure}

\begin{figure}[tbh]
\includegraphics[width=3.20in,clip]{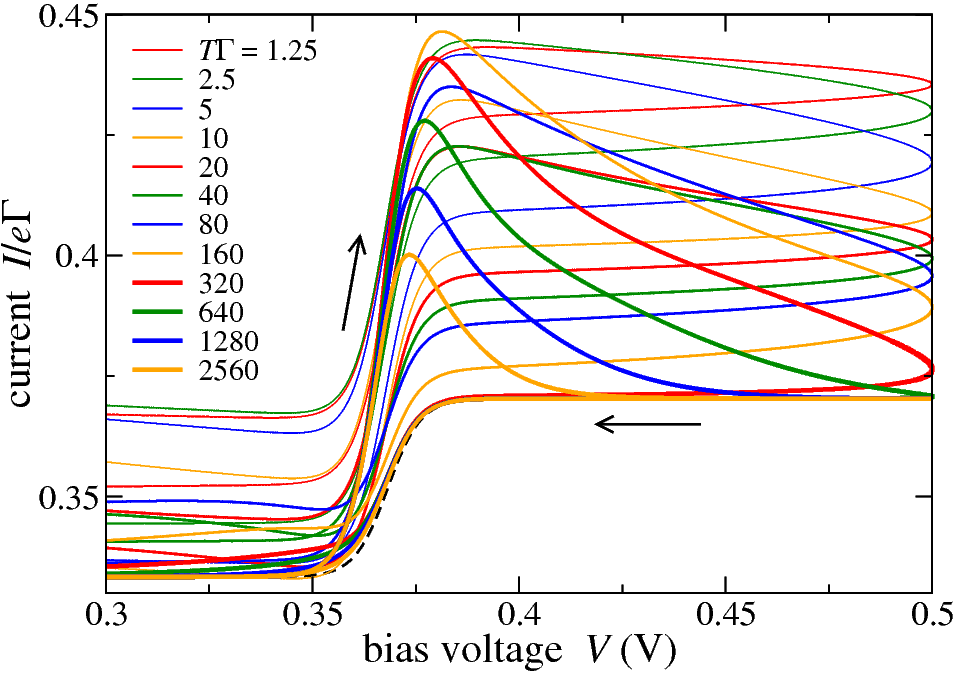}
\caption{\label{fig.per_I}(Color online) Details of the current vs voltage
curves for periods $T=1.25$, \dots, $2560\,\Gamma^{-1}$ from Figs.\
\ref{fig.per_fast}, \ref{fig.per_medium}, and \ref{fig.per_slow}. The
arrows indicate the direction in which the hysteresis loops are traversed.
Black dashed curve: Stationary state.}
\end{figure}

To elucidate the role of the driving period $T$, we have determined the
periodic behavior for various $T$'s as described in Sec.\ \ref{sec.master},
keeping all other parameters fixed. Figures \ref{fig.per_fast},
\ref{fig.per_medium}, and \ref{fig.per_slow} show single hysteresis loops for
current, charge, and spin in the periodic regime. Since the hysteresis in the
current is not very pronounced on the scale of the figures, we show an
enlargement close to the voltage maximum in Fig.\ \ref{fig.per_I}. The natural
time scale is $\Gamma^{-1} = (2\pi|t_\mathrm{hyb}|^2 N_\mathrm{maj})^{-1}$, the
inverse of the characteristic tunneling rate. Note that for a typical current of
$200\,\mathrm{pA}$ in the transmitting regime,\cite{PPL00,HGF06} the tunneling
rate is on the order of $\Gamma \approx 10^9 \,\mathrm{s}^{-1}$.
In the limit of rapid driving ($T\to 0$),
shown in Fig.\ \ref{fig.per_fast}, all hysteresis loops close
since the molecule cannot follow the rapid bias. The charge and the
spin approach constant values determined by the time-averaged rates,
\be
\overline{R}_{n\to m} = \frac{1}{T} \int_0^T dt\,R_{n\to m}(t) .
\ee
The current loop also closes but it does not become a
horizontal line since the current in Eq.\ (\ref{3.current.3}) explicitly
depends on the instantaneous rates, which in turn depend on the instantaneous
bias.

We note that the hysteresis loops for the current, in particular at rapid
driving, show additional steplike features absent from the stationary curves.
These features are due to excited-state-to-excited-state (ETE) transitions. Such
transitions are also observed in the stationary state,\cite{GTT08} but only if
the energy of the ETE transition is higher than the
energy of the transition populating its initial state. This restriction is
relaxed for dynamical measurements, where the initial state can be transiently
occupied even when the voltage is not sufficiently high to
populate it in the stationary regime. Thus, additional spectroscopic information
on ETE transition energies and lifetimes (from the width of the current steps)
can be obtained from dynamical measurements.

In the limit of slow driving, $T\to \infty$, the hysteresis loops must also
close since the observables approach the stationary values.
The current loops in Fig.\ \ref{fig.per_slow}(a)
indeed close for slow driving. In particular, the capacitive response (open
loop at $V=0$) decreases rapidly together with the charge at zero
voltage. However, the charge and the spin in Figs.\ \ref{fig.per_slow}(b) and
\ref{fig.per_slow}(c), respectively, have not approached the stationary
curve even at $T=2560\,\Gamma^{-1}$. The reason for this requires some
discussion.

All quantities show steplike behavior for large $T$ with two pairs of steps at
voltages of about $\pm 0.20\,\mathrm{V}$ and about $\pm 0.37\,\mathrm{V}$.
It is useful to refer to the level scheme in Fig.\ \ref{fig.Evsm} to understand
what happens at these voltages. At $|V|=V_1\equiv
2(E_{1,\pm 5/2}-E_{0,\pm 2}) = 0.2\,\mathrm{V}$, the excess
energy $eV/2$ of electrons appearing in Eqs.\ (\ref{3.rate.1})--(\ref{3.rate.4})
matches the lowest transition energy $E_{1,\pm 5/2}-E_{0,\pm 2}$
out of the ground states with occupation
number $n=0$ and magnetic quantum numbers $m=\pm 2$ to the states $n=1$, $m=\pm
5/2$. These transitions are denoted by solid arrows in Fig.\
\ref{fig.Evsm}. From the excited states $n=1$, $m=\pm 5/2$, the molecule can
only relax back to the ground states by emitting one electron into the leads. No
other transitions are allowed for sequential tunneling. Thus the molecule cannot
overcome the anisotropy barrier and the imbalance
$\Delta M \equiv \langle \mathrm{sgn}(S_\mathrm{tot}^z) \rangle$ between
positive and negative
$m$ cannot be relaxed.\cite{TiE06} In fact, this statement is only rigorously
true at zero temperature. At finite temperatures, there is a thermally activated
process in which the molecule goes from the ground states to a state with
$n=1$, $m=\pm 3/2$ (dashed arrows in Fig.\ \ref{fig.Evsm}), but the transition
rate for this process is exponentially suppressed by the tail of the Fermi
function.

At the second step at $|V|=V_2\equiv
2(E_{1,\pm 3/2}-E_{0,\pm 2}) = 0.367763\,\mathrm{V}$, the excess energy
$eV/2$ matches the transition energy $E_{1,\pm 3/2}-E_{0,\pm 2}$ from the
ground states to a state with $n=1$ and $m=\pm 1/2$ (dashed arrows in Fig.\
\ref{fig.Evsm}). But then the molecule can overcome the anisotropy barrier by a
series of sequential-tunneling transitions that are either exothermal or
endothermal with a transition energy lower than $eV_2/2$. Consequently, for
voltages close to $V_2$, the rate for spin relaxation
over the barrier crosses over from exponentially small
to a sizable fraction of $\Gamma$.\cite{TiE06}
Relaxation over the barrier is still slower than relaxation not crossing the
barrier since it involves several sequential-tunneling transitions, each of
which competes with a transition in the opposite direction.

We can now understand the behavior of the spin for slow driving in Fig.\
\ref{fig.per_slow}: Above the second threshold, $V > V_2$, relaxation over
the barrier is not exponentially suppressed, and the system indeed approaches
the stationary behavior for slow driving. As the voltage is lowered into the
range
$V_1 < V < V_2$, the imbalance $\Delta M$ between positive and negative $m$ is
frozen in. As long as the driving is not exponentially slow, the system will
relax under the constraint of fixed $\Delta M$. In the range $V_1 < V < V_2$
the frozen-in value of $\Delta M$ is close to the stationary value so that the
system can still relax to a state close to the stationary one. When the
voltage falls below $V_1$, all transitions out of the two ground states in
Fig.\ \ref{fig.Evsm} are thermally suppressed, and the system relaxes toward the
ground states with the imbalance $\Delta M$ approximately conserved.

For $-V_2<V<-V_1$, the lowest-energy transitions become active again but now the
frozen value of $\Delta M$ is very different from the stationary
one at negative voltages. The system is still with a high probability on the
left-hand side of the barrier ($\Delta M<0$). The most relevant transition is
the one from the state $n=0$, $m=-2$ to the state $n=1$, $m=-5/2$, which
requires a spin-down electron. However, for negative voltages predominantly
spin-up electrons are injected. The transition rate
for this process is thus suppressed by the spin-polarization factor $p$ in Eq.\
(\ref{3.rate.2}). Consequently, the average occupation $\langle n\rangle$ is
suppressed relative to the stationary case, leading to the plateau at $\langle
n\rangle \approx 0.336$ in Fig.\ \ref{fig.per_slow}(b). Finally, for $V<-V_2$,
spin relaxation over the barrier becomes active again, the imbalance $\Delta M$
is unfrozen, and the system can relax to the stationary state at slow driving.

In summary, the device does approach the stationary regime in the limit
of slow driving, $T\to\infty$, but to reach this regime, the period $T$ has to
be large compared to the \emph{exponentially long} spin relaxation time. We
thus find a separation of time scales: The spin relaxation time is
generically long compared to typical relaxation times for processes not
crossing the anisotropy barrier. This can be compared to the dynamics of a
molecule without local spin but involving a vibrational mode:\cite{DYG12,Koch}
In this case small Franck-Condon matrix elements suppressing low-energy
transitions can lead to a separation of time scales.

\begin{figure}[tbh]
\includegraphics[width=3.10in,clip]{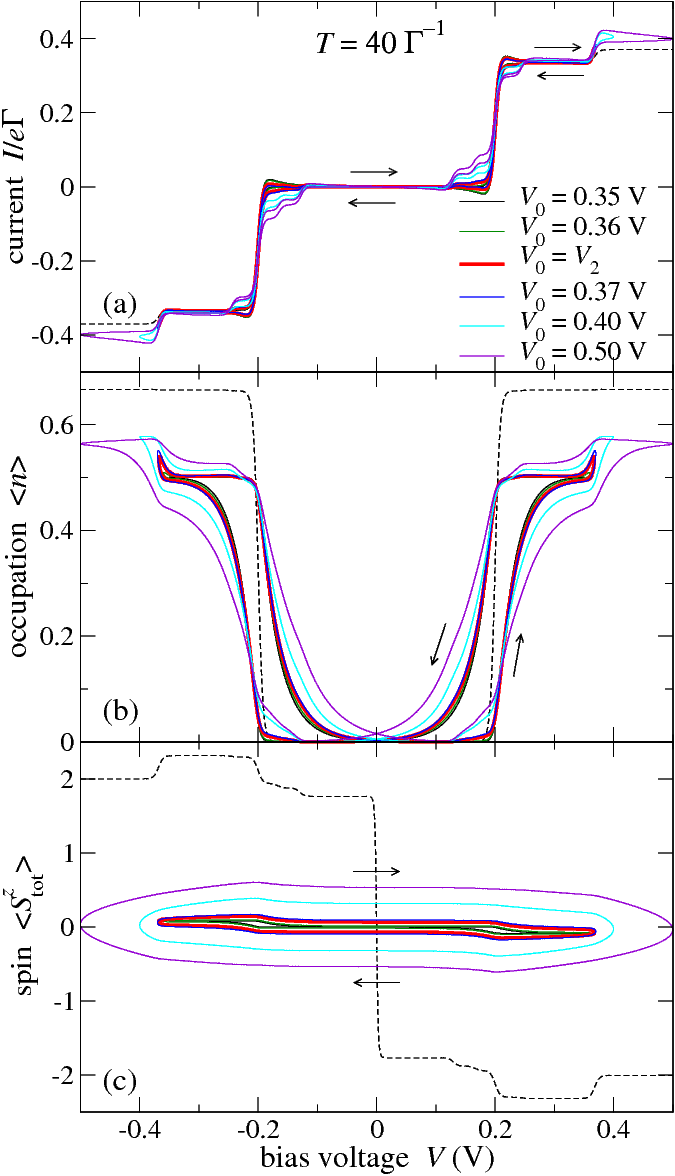}
\caption{\label{fig.per_V0}(Color online) (a) Current, (b) occupation
number, and (c) \textit{z} component of the total spin vs voltage for
voltage amplitudes $V_0 = 0.35\,\mathrm{V}$, $0.36\,\mathrm{V}$,
$0.367763\,\mathrm{V}$ (${=}\:V_2$), $0.37\,\mathrm{V}$, $0.40\,\mathrm{V}$, and
$0.50\,\mathrm{V}$ and period $T=40\,\Gamma^{-1}$. The other parameters are as
in Figs.\ \ref{fig.onecase}--\ref{fig.per_slow}.
A single hysteresis loop in the periodic regime is shown in each case; the loop
at the threshold amplitude, $V_0=V_2$, is highlighted as a heavy red (gray)
curve. The arrows indicate the directions in which the loops are traversed.
The black dashed curves show the voltage dependence in the stationary state.}
\end{figure}

\subsection{Dependence on the voltage amplitude}

Since the imbalance $\Delta M$ between positive and negative magnetic quantum
numbers is effectively frozen in when the voltage drops below the threshold
$V_2$, we expect the behavior of the device to change dramatically when the
amplitude $V_0$ of the bias $V(t) = V_0\, \sin \omega t$ is tuned
through $V_2$. To exhibit this dependence, we have determined the
periodic behavior for amplitudes $V_0$ through the threshold $V_2$, keeping all
other parameters fixed. Figure \ref{fig.per_V0} shows single hysteresis loops
for current, charge, and spin in the periodic regime. As expected, the spin
hysteresis loops are nearly closed for voltage amplitudes $V_0$ below the
threshold $V_2$ and open up above the threshold. Then the system
can overcome the barrier for part of the driving period so that the spins
injected from the magnetized leads can reverse the local spin.

\begin{figure}[tbh]
\includegraphics[width=3.10in,clip]{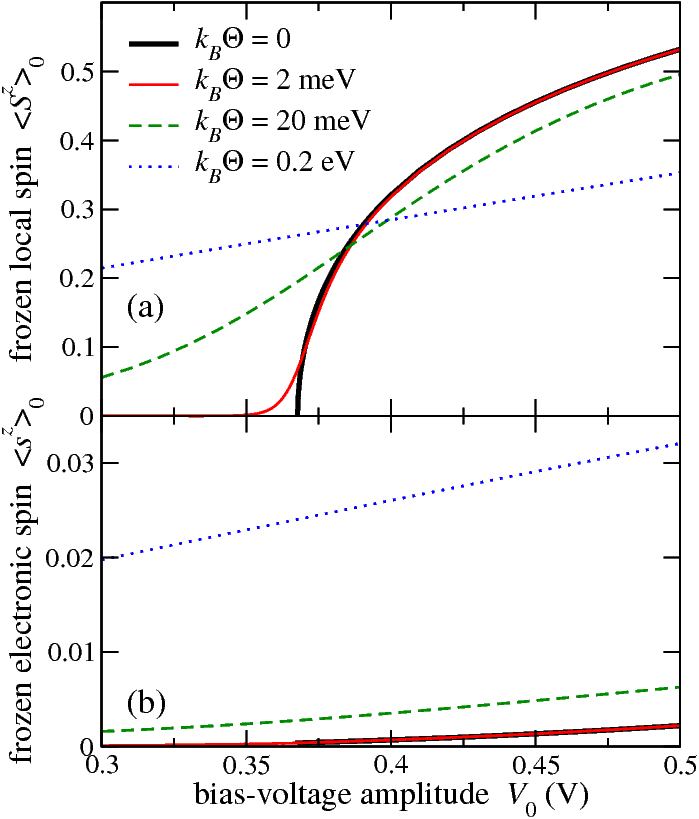}
\caption{\label{fig.scrit_V0}(Color online) (a) Frozen local spin $\langle
S^z\rangle_0$ and (b) frozen electronic spin $\langle s^z\rangle_0$ at $V(t)=0$
and $V'(t)>0$ vs voltage amplitude at thermal energies $k_B\Theta = 0$,
$2\,\mathrm{meV}$, $20\,\mathrm{meV}$, and $0.2\,\mathrm{eV}$.
The other parameters are as in Fig.\ \ref{fig.per_V0}.}
\end{figure}

A reasonable measure of the size of the spin hysteresis loop---which represents
the effectiveness of the device in storing information---is the frozen spin at
$V=0$ with $V'(t)>0$. We plot
the local and the electronic contributions to the frozen spin as functions of
the voltage amplitude for zero and finite temperatures in Fig.\
\ref{fig.scrit_V0}. To obtain the values at zero temperature, we proceed
differently from the method outlined in Sec.\ \ref{sec.master}: For
$k_B\Theta=0$, the Fermi function becomes a step function and thus the
transition rates in Eqs.\ (\ref{3.rate.1})--(\ref{3.rate.4}) are piecewise
constant functions of the instantaneous voltage $V(t)$ and, therefore, of
time. This allows us to obtain the stochastic matrix $\Pi$
analytically as a product of a finite number of matrices describing the time
evolution over time intervals with constant rates.

The frozen local spin $\langle S^z\rangle_0$ in Fig.\ \ref{fig.scrit_V0}(a)
shows critical behavior for vanishing temperature. Taking the square
we find that the critical exponent is $1/2$,
$\langle S^z\rangle_0 \sim (V_0-V_2)^{1/2}$.
The singularity is smeared out at finite temperatures. For the frozen
electronic spin, critical behavior is not evident.

The origin of the exponent $1/2$ is that the fraction of time during which the
voltage is large enough to overcome the barrier scales with $(V_0 -
V_2)^{1/2}$ {provided that} $V(t)$ is analytic close to its extrema. Taking, for
example, the first voltage maximum, the end points of this time interval
are obtained by solving $V(T/4\pm \Delta t/2) = V_2$.
If $V(t)$ is analytic, we can expand around the maximum,
$V_0 + V''(T/4)\Delta t^2/8 = V_2$,
the solution of which gives $\Delta t \sim (V_0 - V_2)^{1/2}$. We now expand
the stochastic matrix $\Pi$ for small $\Delta t$ and use perturbation theory
to find the probability vector for the periodic state satisfying
$\mathbf{P}^\mathrm{per} = \Pi\, \mathbf{P}^\mathrm{per}$. The leading
correction is linear in $\Delta t$ and thus proportional to $(V_0 - V_2)^{1/2}$.

We can therefore draw two conclusions: (i) Quite generally, all observables
should inherit a
$(V_0 - V_2)^{1/2}$ correction from $\mathbf{P}^\mathrm{per}$ above the
threshold. (ii) The same argument also applies whenever additional transitions
become energetically allowed at higher voltage amplitudes. Thus there
should be corresponding singular terms associated with these transitions. We
have checked that this is borne out by the numerical results.

\begin{figure}[tbh]
\includegraphics[width=3.20in,clip]{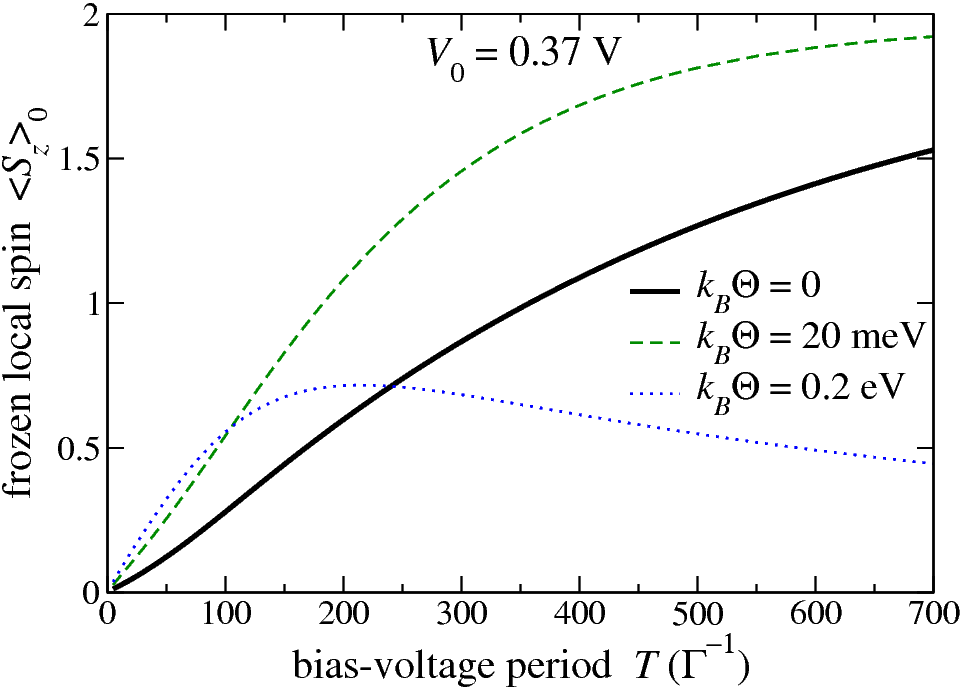}
\caption{\label{fig.scrit_T}(Color online) Frozen local spin $\langle
S^z\rangle_0$ vs voltage period at thermal energies $k_B\Theta = 0$,
$20\,\mathrm{meV}$, and $0.2\,\mathrm{eV}$. The voltage amplitude is
$V_0=0.37\,\mathrm{V}$, slightly above the threshold $V_2$. The other parameters
are as in Fig.\ \ref{fig.per_V0}.}
\end{figure}

We now turn to the dependence of the frozen spin on the driving frequency.
Figure \ref{fig.scrit_T} shows the frozen local spin $\langle S^z\rangle_0$ as a
function of the voltage period $T$ for three temperature values. The voltage
amplitude $V_0$ is slightly above the threshold $V_2$ for spin relaxation. At
rapid driving, $T\to 0$, the frozen spin goes to zero, as expected since the
system cannot follow the rapidly changing bias. For very large periods,
the frozen spin should approach the stationary value at zero bias, which is also
zero. For the (unphysically) high temperature $k_B\Theta = 0.2\,\mathrm{eV}$,
Fig.\ \ref{fig.scrit_T} indeed shows this behavior. On the other hand, at zero
temperature, the frozen spin \emph{never} approaches zero for large $T$ since
there is
no spin relaxation for $|V(t)|<V_2$ so that the system can never reach the
stationary state with zero average spin at zero bias.

\begin{figure}[tbh]
\includegraphics[width=3.10in,clip]{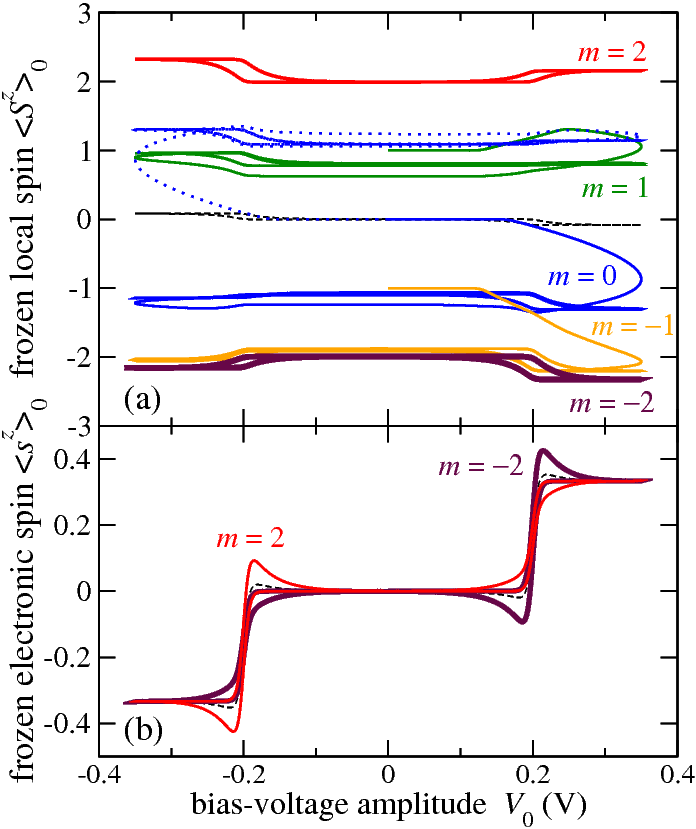}
\caption{\label{fig.nIquasi}(Color online) (a) Solid curves: Five periods of
the \textit{z} component of the total spin, $\langle S^z_\mathrm{tot}\rangle$,
vs voltage for different initial pure states with occupation number $n=0$
and magnetic quantum
numbers $m=2$, $1$, $0$, $-1$, and $-2$. The voltage amplitude is
$V_0=0.35\,\mathrm{V} < V_2$ and the voltage period is $T=40\,\Gamma^{-1}$. The
other parameters are as in Figs.\ \ref{fig.onecase}--\ref{fig.per_slow}.
Dotted curve: The same for the
initial state with $m=0$ but for the voltage shifted in time by half a
period. (b) Solid curves: Five periods of the current for the two cases
with $m=2$ and $m=-2$. The parameters are the same as in (a).
The black dashed curves show the true periodic state that
the system would reach after an exponentially long time.}
\end{figure}

\subsection{Quasiperiodic regime}

As we have seen, the anisotropy barrier leads to a separation of time scales
for relaxation over the barrier and relaxation staying on one side of the
barrier. It thus makes sense to consider the intermediate regime reached after
the fast transients have died out, but before the slow relaxation has become
effective. We call this the ``quasiperiodic'' regime.

As discussed above, the
true periodic state is unique since it is determined by the stationary solution
of an ergodic discrete-time Markov process. Thus the system will eventually
converge to the same periodic solution regardless of its initial state. However,
the
same is not true for the quasiperiodic state, which will depend on the initial
condition and, in principle, also on the protocol with which the bias is
switched on. This is illustrated in Fig.\ \ref{fig.nIquasi}(a), which shows the
first five periods of the time evolution of the total spin $\langle
S^z_\mathrm{tot}\rangle$ starting from different initial states. The voltage
amplitude is $V_0 = 0.35\,\mathrm{V}<V_2$ so that the relaxation over the
barrier is exponentially slow. After a few periods, a quasiperiodic
regime is reached, which indeed depends on the initial state. After a much
longer time, during which the system can relax over the barrier, all curves
would approach the true periodic hysteresis loop (dashed curve).

We have also verified that the quasiperiodic loop does depend on the way the
voltage is switched on. The dotted blue
(gray) curve in Fig.\ \ref{fig.nIquasi}(a) shows the spin for the same initial
state as the solid blue (gray) curve; the only difference is that the
voltage is shifted in time by half a period, i.e., $V(t) = -V_0\sin \omega t$.
Since the system starts in the state with
$m=0$, which is the one on top of the anisotropy barrier in Fig.\
\ref{fig.Evsm}, the sign of the voltage applied during the first half period
determines the dominant spin direction of the injected electrons. This
determines the probabilities for the system to relax into states with positive
or negative $m$. The resulting imbalance $\Delta M$ is then frozen in because
the voltage amplitude is below the threshold $V_2$. Consequently, the sign of
$V(t)$ during the first half period determines the spin polarization.

\begin{figure}[tbh]
\includegraphics[width=3.20in,clip]{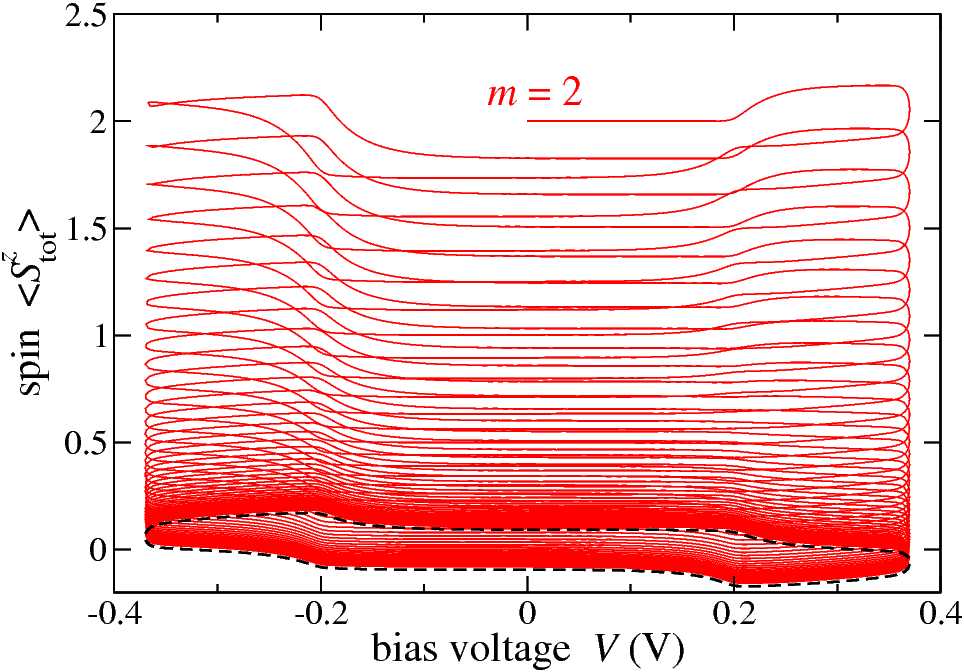}
\caption{\label{fig.nquasi2}(Color online) Solid red (gray) curve: $50$ periods
of the \textit{z} component of the total spin vs voltage for the initial pure
state with occupation number $n=0$ and
magnetic quantum number $m=2$. The voltage amplitude is
$V_0=0.37\,\mathrm{V} > V_2$ and the voltage period is $T=40\,\Gamma^{-1}$. The
other parameters are as in Fig.\ \ref{fig.nIquasi}. Dashed black curve:
Periodic state.}
\end{figure}

It is illuminating to compare the behavior when the voltage amplitude is above
the threshold $V_2$. In this case, we do not expect a separation of time scales.
Figure \ref{fig.nquasi2} indeed shows that the spin approaches the periodic
hysteresis loop without reaching any intermediate quasiperiodic
regime. The relaxation over the barrier is still slow since it involves several
sequential-tunneling transitions and is only active for a small fraction of the
time.

\begin{figure}[tbh]
\includegraphics[width=3.30in,clip]{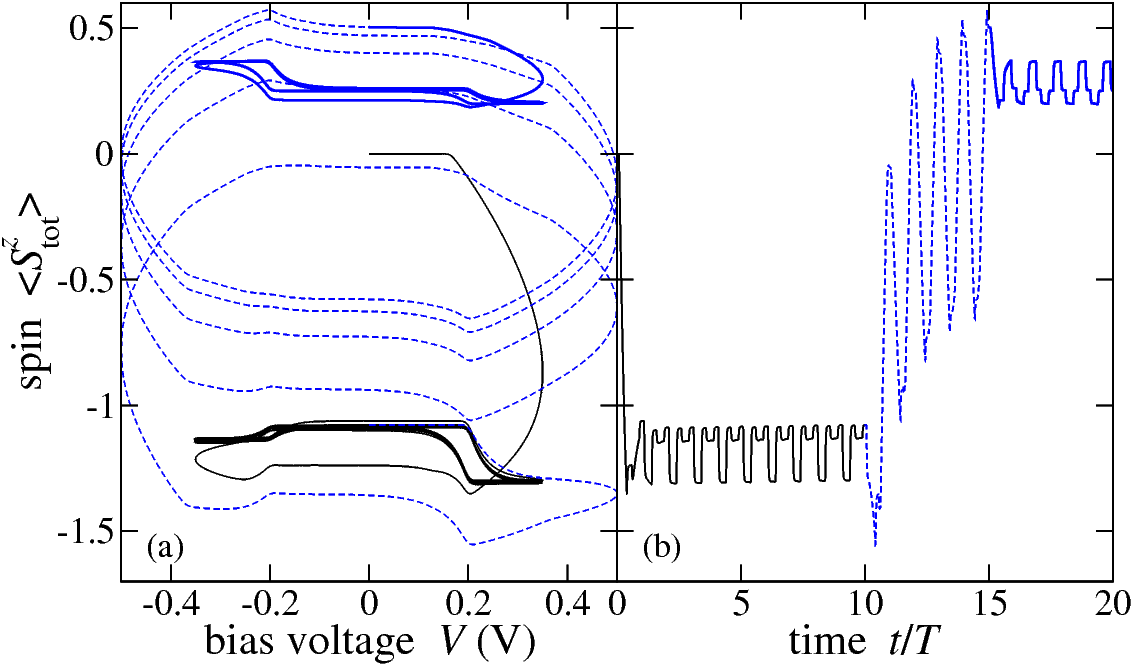}
\caption{\label{fig.nswitch}(Color online) \textit{z} component of the total
spin vs (a) voltage and (b) time, for a process
starting from the pure state with $n=0$ and $m=0$ and consisting of ten periods
at the amplitude $V_0=0.35\,\mathrm{V}<V_2$, five periods at
$V_0=0.5\,\mathrm{V}>V_2$,
and five periods at $V_0=0.35\,\mathrm{V}<V_2$. The voltage period is held
fixed to $T=40\,\Gamma^{-1}$. The other parameters are as in the previous
figures.}
\end{figure}

The spin polarization in the intermediate, quasiperiodic regime can be
\emph{read out} by a transport experiment: Figure \ref{fig.nIquasi}(b)
shows the current hysteresis loops for the two cases with initial value $m=\pm
2$ (solid curves) and also the true periodic behavior (dashed curve). The loops
are clearly different. To use our model
system as a memory device, we also need a protocol for \emph{writing} the spin.
This is possible by increasing the
voltage amplitude over the threshold, and making the period $T$
sufficiently long. Then the spin approaches a large hysteresis loop on a
time scale of a few periods. Reducing the voltage amplitude below
the threshold while the spin is large in magnitude freezes the imbalance
$\Delta M$. For illustration,
we plot in Fig.\ \ref{fig.nswitch} the spin for a process starting from the pure
state with $n=0$ and $m=0$ and consisting of ten periods at a small voltage
amplitude $V_0<V_2$, followed by five periods at $V_0>V_2$,
and eventually by five more periods at the smaller amplitude.
Figure \ref{fig.nswitch} shows the spin as a function of bias and of time. This
protocol clearly switches the system between two distinct quasiperiodic
states. If we had reduced the amplitude half of a period earlier or later, i.e.,
when $\langle S^z_\mathrm{tot}\rangle$ was negative, a negative spin would have
been written.

The typical time scale of the switching
in Fig.\ \ref{fig.nswitch} is a few times $T=40\,\Gamma^{-1} \sim 4\times
10^{-8}\,\mathrm{s}$. This should be compared to the switching time of
memristive systems containing a MgO layer between ferromagnetic electrodes,
which is on the order of seconds, probably because it involves the
displacement of oxygen vacancies.\cite{HMB08,KKR09} The switching mechanism
proposed in the present paper can be much faster since it involves neither
nuclear motion, as Refs.\ \onlinecite{HMB08} and \onlinecite{KKR09} and also
the spin transition
considered by Miyamachi \textit{et al.}\cite{MGD12} do, nor the motion of
domain walls, which is the mechanism studied by M\"unchenberger \textit{et
al.}\cite{MRT12}

\section{Summary and conclusions}
\label{sec.summary}

Memristive (memory resistive) properties of a SMM weakly
coupled to two ferromagnetic leads have been investigated. To that end, we have
studied three observables: the average current through the molecule, the
occupation (charge), and the \textit{z} component of the total spin on the
molecule. We have obtained the hysteresis loops for these quantities vs the
applied bias. The device is not a purely memristive system, as evidenced by the
open hysteresis loop for the current. Instead, the transient charging of the
molecule leads to a partially capacitive response. This capacitive response is
governed by the fast charge relaxation. The device combines memristive with
spintronics functionality facilitated by a large polarization of the molecular
spin.

For rapid driving, i.e., for driving frequencies on the order of the
characteristic tunneling rate $\Gamma$, the memory dependence is suppressed and
the hysteresis loops close. The
current-voltage characteristics nevertheless show additional spectroscopic
features not seen in the stationary (dc) current-voltage curve. In the opposite
limit of very slow driving, memory effects are also suppressed since the system
eventually approaches the stationary behavior. However, the period of the
voltage required to reach this stationary regime can be exceedingly long in the
presence of easy-axis anisotropy. The origin
of this dynamical slowing down is that relaxation of the
molecular spin over the anisotropy barrier becomes suppressed below a certain
bias voltage so that any imbalance between positive and negative spin
polarization is essentially frozen in. It is the very slow relaxation over
the barrier that governs the eventual closing of the hysteresis loops.

At zero temperature, the frozen occupation number and spin at zero bias
exhibit non-analyticities when the voltage amplitude crosses the threshold
for relaxation over the barrier. The
non-analytic contributions scale with the voltage distance from the transition
point with a critical exponent of $1/2$. The singular behavior is smeared out at
finite temperatures.

We have seen that the easy-axis anisotropy naturally leads to a separation of
time scales: If the bias voltage is below a critical threshold, relaxation over
the barrier is exponentially suppressed compared to relaxation between states
on the same side of the barrier, by the tail of the Fermi function appearing in
the sequential-tunneling rates.
Now, the question arises as to whether any processes
neglected in our approach become relevant in this regime. Sequential tunneling
is of the order of $t_\mathrm{hyb}^2 \propto \Gamma$. At the order
$t_\mathrm{hyb}^4 \propto \Gamma^2$, cotunneling occurs: An electron can tunnel
coherently from one lead to the other. During this process, it can flip its
spin, which leads to a transition of the molecular state without a change of the
occupation number, but with a change of the spin by
one unit. Out of the ground states, this process becomes active when $eV$
matches the energy difference $eV_\mathrm{cot} \equiv
E_{0,\pm 1}-E_{0,\pm 2}$ between the states with $m=\pm 1$ and $m=\pm 2$,
which for the parameters chosen here happens at a voltage below the
Coulomb-blockade threshold $V_1$, see Fig.\ \ref{fig.Evsm}.
(Note that the full bias $eV$ enters in this case.)
Beyond the voltage $V_\mathrm{cot}$, the system can overcome the anisotropy
barrier by
cotunneling. Thus there is a regime where relaxation over the barrier is
dominated by cotunneling, which is down by a factor of $\Gamma$ compared to
sequential tunneling, but does not involve an exponentially small factor at low
temperatures. At smaller voltages, $|V|<V_\mathrm{cot}$, cotunneling is
also exponentially suppressed and processes of even higher order in $\Gamma$
become important. Eventually, the direct transition from one ground state
to the other involving a change of the spin by four units occurs at order
$\Gamma^8$. This process happens even at zero bias and is thus never suppressed
by exponential factors.

Still, below the threshold $V_2$ for spin relaxation due to sequential
tunneling, the relaxation rate is at least suppressed by a factor of $\Gamma$.
Thus, for sufficiently weak coupling between the molecule and the leads, there
is still a wide separation of time scales.
We finally note that transverse spin-anisotropy terms, or a transverse magnetic
field, can lead to spin tunneling through the barrier and thereby open another
channel for spin relaxation.

The separation of time scales opens up an intermediate time regime where fast
transients have died out, but relaxation over the barrier has yet to become
effective. In this regime, a quasiperiodic dependence of all observables on
the bias is observed. Unlike the true periodic state, the quasiperiodic
hysteresis loop depends on the initial conditions and the protocol by which
the bias is switched on. In view of
the long lifetime that can be realized experimentally, this means that
this property can be used to store information. Indeed, we have demonstrated
that this
spin information can be read out by measuring the alternating current, and that
it can be rewritten at will by judiciously changing the voltage amplitude.
Note that we have discussed a two-terminal device. None of this functionality
requires a gate electrode (although the latter would add extra flexibility),
which should make the practical implementation significantly easier.

Finally, it is also worth noting that the presence of different time scales and
long relaxation times makes molecular magnets ideally suited for a host of
neuromorphic applications,\cite{pershin12a} ranging from learning
circuits\cite{pershin09b} and associative memory\cite{pershin10c} to the
massively parallel solution of optimization problems.\cite{pershin11d}
While neuromorphic and memristive devices based on
magnetic solid-state structures have been
suggested,\cite{HMB08,KKR09,MRT12,SAP12}
molecular magnets potentially offer higher integration densities, faster
switching, and lower power consumption. These features make them ideal
candidates for neuromorphic computing.


\acknowledgments

We would like to thank T. Ludwig and M. P. Sarachik for useful
discussions. Financial support by the Deutsche Forschungsgemeinschaft, in part
through Research Unit 1154, \textit{Towards Molecular Spintronics}, and
by the US National Science Foundation Grant No.\ DMR-0802830 is gratefully
acknowledged.

\end{document}